\documentclass[12pt,preprint]{aastex}

\newcommand\msun{\hbox{\,M$_\odot$}}
\newcommand\rsun{\hbox{\,R$_\odot$}}

\newcommand\kms{ km~s$^{-1}$}

\newcommand\betlyr{$\beta$~Lyr~}
\def\kms{\ifmmode{\rm km\thinspace s^{-1}}\else km\thinspace s$^{-1}$\fi}

\shorttitle{Imaging \betlyr}
\shortauthors{Zhao et al.}

\begin{document}

\title{First Resolved Images of the Eclipsing and Interacting Binary $\beta$ Lyrae }

\author{M.~Zhao\altaffilmark{1},
D. Gies\altaffilmark{2},
J.~D.~Monnier\altaffilmark{1},
N. Thureau\altaffilmark{3},
E. Pedretti\altaffilmark{3},
F. Baron\altaffilmark{4},
A. Merand\altaffilmark{2},
{T.~ten Brummelaar}\altaffilmark{2},
H. McAlister\altaffilmark{2},
S. T. Ridgway\altaffilmark{5},
N. Turner\altaffilmark{2},
J. Sturmann\altaffilmark{2},
L. Sturmann\altaffilmark{2},
C. Farrington\altaffilmark{2},
P. J. Goldfinger\altaffilmark{2}
}

\altaffiltext{1}{mingzhao@umich.edu: University of Michigan Astronomy Department,
941 Dennison Bldg, Ann Arbor, MI 48109-1090, USA}
\altaffiltext{2}{The CHARA Array, Georgia State University, USA}
\altaffiltext{3}{University of St. Andrews, Scotland, UK}
\altaffiltext{4}{University of Cambridge, UK}
\altaffiltext{5}{National Optical Astronomy Observatory, NOAO, Tucson, AZ}

\begin{abstract}
We present the first resolved images of the eclipsing  binary $\beta$ Lyrae, obtained with the CHARA Array interferometer and the MIRC combiner in the $H$ band. The images clearly show the mass donor and the thick disk surrounding the mass gainer at all six epochs of observation. The donor is brighter and generally appears elongated in the images, the first direct detection of photospheric tidal distortion due to Roche-lobe filling.  We also confirm expectations that the disk component is more elongated than the donor and is relatively fainter at this wavelength.
Image analysis and model fitting for each epoch were used for calculating the first astrometric orbital solution for $\beta$~Lyrae, yielding precise values for the orbital inclination and position angle.
The derived semi-major axis  also allows us to estimate the distance of $\beta$ Lyrae; however,
systematic differences between the models and the images limit the accuracy of our distance estimate to about 15\%. To address these issues, we will need a more physical, self-consistent model to account for all epochs as well as the multi-wavelength information from the eclipsing light curves.

\end{abstract}


\section{Introduction}

Interacting binaries are unique testbeds for many important astrophysical processes, such as mass and momentum transfer, accretion, tidal interaction, etc. These processes provide information on the evolution and properties of many types of objects, including low-mass black holes and neutron stars (in low-mass X-ray binaries), symbiotic binaries, cataclysmic variables, novae, etc.  Although these types of objects are widely studied by indirect methods such as spectroscopy, radial velocity, and sometimes eclipse mapping,  very few of them have been directly resolved because they are very close to each other and far away from us.
Thus, directly imaging  interacting binaries, although very challenging, will greatly help us to improve our understanding of these objects.

The star $\beta$ Lyrae (Sheliak, HD 174638, HR 7106, $V$ = 3.52, $H$=3.35) is a well known interacting and eclipsing binary that has been widely studied since its discovery in 1784 \citep{Goodricke1785}.  According to the current picture \citep{Harmanec2002}, the system consists of a B6-8~II Roche-lobe filling mass-losing star, which is generally denoted as the donor or the primary,  and an early B type mass-gaining star that is  generally denoted as the gainer or the secondary. The donor, which was initially more massive than the gainer, has a current mass of about 3 \msun, while the gainer has a mass of about 13 \msun. It is thought that the gainer is completely embedded in a thick accretion disk with bipolar jet-like structures perpendicular to the disk, which creates a light-scattering halo above its poles. The orbit of the system is highly circular \citep{Harmanec1993}, and is very close to edge-on \citep{Linnell2000}. Recent RV study on the  ephemeris of the system gives a period of $12.^d94$ \citep{Ak2007}. The period is increasing at a rate of $\sim19$ sec  per year due to the high mass transfer rate, $2\times10^{-5}\msun$ yr$^{-1}$, of the system. 
%In addition to the orbital period, a 283-day periodicity also exists in the minimum of the primary eclipse, but its cause is still an open question \citep[e.g.,][]{Kreiner1999}. 

The primary eclipse of the light curve (i.e., at phase 0) corresponds to the eclipse of the donor. In the $UBV$ bands, the surface of the donor is brighter than that of the gainer, and  therefore the primary minimum is deeper than the secondary minimum. At longer wavelengths, however, the studies of  \citet{Jameson1976} and \citet{Zeilik1982} suggest that the relative depth of the secondary minimum in the light curve gradually deepens and becomes deeper than the primary minimum at wavelengths longer than  $3.6\mu m$.

%Despite intensive study, \betlyr is still enigmatic with many questions left to be addressed. Direct imaging will undoubtedly shed light  on our understanding of this system and its circumstellar properties. 

%Direct imaging will undoubtedly shed light  on our understanding of this system and its circumstellar properties, however 

Light curve studies and theoretical models have shown that, at the distance of 296pc \citep{van-Leeuwen2007}, the estimated separation of the binary is only 0.92 milli-arcsecond (hereafter $mas$, $58.5\rsun$). The angular diameter of the donor is  $\sim$0.46 mas (29.4\rsun), and the disk surrounding the gainer is only $\sim1$ mas across \citep[e.g.,][]{Linnell2000, Harmanec2002}. The goal of directly imaging $\beta$ Lyr, therefore,  requires the angular resolution only achievable by today's long-baseline interferometers.   
Recently, \citet{Schmitt2008} used the NPOI interferometer to image 
successfully the H$\alpha$ emission of $\beta$ Lyr, an update to the pioneering work of \citet{Harmanec1996}.  Also, radio work using MERLIN found a nebula  surrounding the secondary but could not resolve its bipolar shape \citep{Umana2000}. Despite recent progress, the individual objects of the system have not been resolved yet, putting even a simple astrometric orbit beyond our reach.

In this study, we present the first resolved images of the \betlyr system at multiple phases, obtained with the CHARA Array and the MIRC combiner. 
We give a brief introduction to our observations and data reduction in \S\ref{observations}. We present our aperture synthesis images with simple models in \S\ref{imaging}. In \S\ref{orbit} we discuss our astrometric orbit of \betlyr and we give the outlook for future work in \S\ref{summary}.

\section{Observations and data reduction}
\label{observations}

Our observations were conducted at the Georgia State University (GSU) Center
 for High Angular Resolution Astronomy (CHARA) interferometer array  
 along with the MIRC instrument. The
 CHARA array, located on Mt. Wilson and consisting of six 1-meter telescopes, is the longest optical/IR 
 interferometer array in the world \citep{Brummelaar2005}.  It has 15 baselines ranging from 34m to 331m, providing resolutions up to $\sim$0.5 mas at H band and $\sim0.7$ mas at K band.
  
\begin{figure}[thb]
\begin{center}
{
\includegraphics[angle=90,width=2.7in]{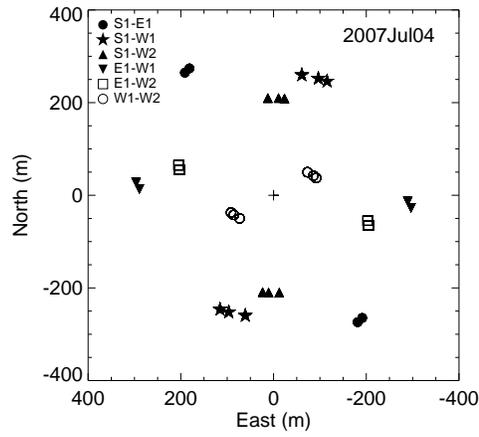}}
\hphantom{.....}
\caption{ % \scriptsize
Telescope spatial coverage of \betlyr on UT 2007Jul04, using the W1-W2-S1-E1 configuration of CHARA. The symbols stand for different baselines. The longest projected baseline in this observation is 328.5m, corresponding to a resolution of 0.52 milli-arcseconds in the $H$ band. The actual UV coverage is similar to this spatial coverage but each point spreads over 8 wavelength channels.
\label{uv}}
\end{center}
\end{figure}

The Michigan Infra-Red Combiner (MIRC) was used here to combine  4 CHARA telescopes together for true interferometric imaging in H band, providing 6 visibilities, 4 closure phases and 4 triple amplitudes simultaneously in 8~narrow spectral bands \citep[see][for details]{Monnier2004,Monnier2006}.  Specifically, the \betlyr system was observed on 6 nights in 2006 and 2007 using an array configuration
optimized for good imaging (equal Fourier coverage in all directions) and following standard observing
procedures \citep[][M. Zhao et al. 2008, in preparation]{Monnier2007sci}. A typical baseline coverage of our observations is shown in Figure     \ref{uv}.
%
%We observed the \betlyr system on 6 nights in 2006 and 2007 using a low resolution spectrograph %which splits the H band into 8 channels,  following the observation procedures discussed in \citet{Monnier2007sci} and \citet{zhao2008b}. 
In short, we observed our target along with two or three calibrators on each night and a complete observing log is listed in Table~\ref{obslog}.

%spread over the whole night. For the purpose of bias subtraction and flux calibration, each set of fringe data is bracketed with measurements of background (i.e., data taken with all beams closed),  shutter suences (i.e., data taken with only one beam open at a time to estimate the amount of light coming from each beam), and foreground (i.e., data taken with all beams open but without fringes). Each object is observed for multiple sets. 
%In order to track the flux variation on each beam in ``real time" to improve the visibility calibration, we  use spinning choppers to temporally modulate the light going into each fiber simultaneously with the fringe measurements. 
%The chopper speeds were set to 25Hz, 30Hz, 35Hz and 40Hz in 2006 and were increased to 55Hz, 65Hz, 75Hz and 85Hz in 2007 to avoid overlapping of chopping frequencies caused by the choppers' intrinsic fluctuation. 
%The log of our observations is listed in Table \ref{obslog}. 
%Figure \ref{uv} shows the baseline coverage of one typical observation on UT 2007Jul04.

The data reduction process follows the pipeline outlined by \citet{Monnier2007sci}.
%, which was validated using 
%data on the binary $\iota$ Peg. 
In brief, after frame co-adding, background subtraction and Fourier transform of the raw data, fringe amplitudes and 
phases are used to form squared-visibilities and triple products.  Photometric calibrations are estimated
using shutter matrix measurements and partial beam chopping.  
%spectrum after bias subtraction. After the fiber coupling 
%efficiencies are estimated using either the chopping signal or 
%direct fit to the fiber profiles,  we obtain 
%uncalibrated squared-visibilities and complex triple amplitudes. 
Finally, calibrators with known sizes  (see Table~\ref{obslog}) are used to calibrate the drifts
in overall system response before obtaining final calibrated squared-visibilites and complex triple amplitudes. 

%===================================

%=====================================

\section{Synthesis imaging and modeling}
\label{imaging}
\begin{figure}[thb]
\begin{center}
{
\includegraphics[angle=0,width=3.45in]{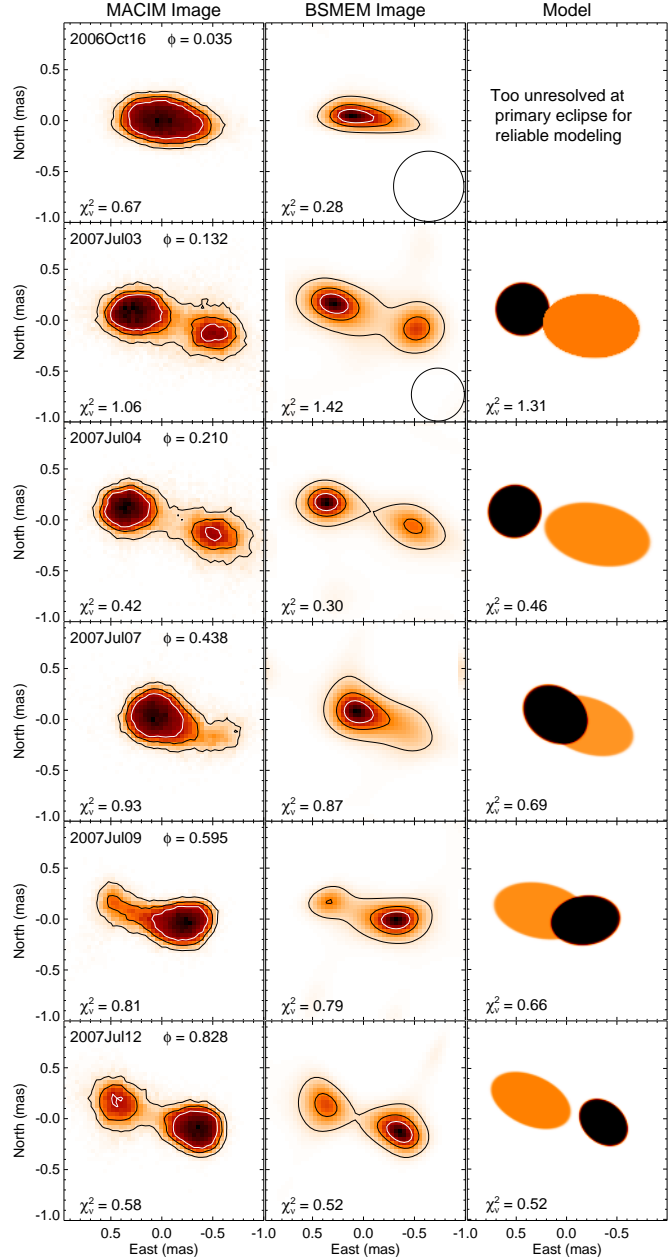}}
\hphantom{.....}
\caption{ % \scriptsize
Reconstructed images and two-component models of \betlyr. The left, middle and right columns show the MACIM, BSMEM and model images respectively. 
Darker colors indicate higher intensity. The darker component is the donor. The contours in the images correspond to 0.3, 0.6, 0.9 of the peak intensity. Observing dates and corresponding phases (from the ephemeris in Ak et al.\ 2007) are labeled in the first column. The best-fit  $\chi^2/DOF$ of each image is labeled in the bottom left corner. The resolution of the reconstructed images is 0.69 mas for the first epoch and 0.52 mas for the other 5 epochs, and the corresponding beams are shown in the first and second epochs in the middle panels respectively. Due to lack of enough resolution and the complexity of the radiative transfer at the first epoch when the star is behind the disk, no reliable model is available for our limited data.
\label{images}}
\end{center}
\end{figure}

For imaging with optical interferometry data, we employed two independent applications: 
``Markov-Chain Imager for Optical Interferometry (MACIM)" \citep{Ireland2006} and
the Maximum-Entropy-based BSMEM \citep{Buscher1994}.  Further description and a detailed comparison of these  algorithms on simulated data appear in \citet{Lawson2004, Lawson2006}.
Both of these algorithms benefit from use of prior information, generally based on lower resolution data.
For $\beta$~Lyr we began each image reconstruction with a two-component Gaussian model which
mainly acts to limit the field-of-view of the image.  The final images do not resemble the priors except in general extent, i.e., the final positions, relative sizes, and relative brightnesses are not 
dependent on the priors.  The final reconstructed images from both methods are shown in Figure \ref{images}.   The MACIM and BSMEM images are consistent with each other, although they use very different algorithms, giving confidence to the image fidelity.   Any differences, such as the
more Gaussian shapes for BSMEM compared to the more ``flat-top" profiles for MACIM, illustrate the limitations of our dataset.   We present here results from both algorithms in lieu of image ``error bars" which are notoriously difficult to define in aperture synthesis imaging.

The six epochs span all phases of the orbit, changing from middle primary eclipse (phase=0.035) to nearly maximum elongation (phase=0.210 and 0.828), and  secondary eclipse (phase=0.438).  
The system is well resolved into two separate components at phases close to the maximum elongation. Since the primary eclipse
is still the deeper one in the $H$-band (Jameson \& Longmore 1976), we can conclude that the object with higher surface brightness is
the mass donor star (i.e., the component moving from left to right in the 2007 July sequence). 
 The donor is partially resolved and appears elongated at all epochs except at phase 0.035 when it is blocked by the disk, directly confirming its Roche-lobe filling picture. The thick disk surrounding the gainer is also resolved and appears elongated. At the first epoch (phase = 0.035),  we see mostly the emission from the disk superposed with a small amount of light from the poles of the donor. 
 %The apparent size of the disk at this epoch is close to 1 mas from the images, similar to estimates from theoretical models \citep{Linnell2000} and the H$\alpha$ disk size of \citet{Schmitt2008}.  
%At the first epoch (phase = 0.035), since most of the light from the donor is blocked, we see mostly the emission from the disk superposed with a small amount of light from the poles of the donor. The apparent size of the disk at this epoch is close to 1 mas, similar to estimates from theoretical models \citep[e.g.,][]{Linnell2000}.  The jet-like structure found in optical studies \citep[etc.]{Harmanec1996, Ak2007} is not seen in the images at any phases,  suggesting that it is too faint in the $H$ band and only appears at certain wavelengths in the optical/$UV$. 

We can extract further information by constructing a simple two-component model to determine the separation and position angle for each epoch. We assume the donor and gainer can be modeled as uniform ellipses.
% lets leave this out unless a referee wants 
Other models, such as two truncated gaussian ellipses, a rain-drop-shaped Roche-lobe filling star with a truncated gaussian disk, etc., were also considered and gave equivalent results due to limited resolution. Therefore, for simplicity and to minimize the
degrees of freedom of the model, uniform ellipses are adopted.
The free parameters in the models are: the semi-major and semi-minor axes of the two components, their individual position angles, the separation and position angle of the system, and the flux ratio of the donor and the disk.
We used ephemeris data from \citet{Ak2007} to fix which component was in front during modeling.
Due to degeneracies in the separation and the dimensions of the blocked component when the  two  are overlapping with each other,  the size of the blocked component is fixed to the average from the two separated epochs,  2007Jul04 and 2007Jul12. 
The best-fit models for all epochs are presented in the third column of Figure \ref{images}
and the resulting positions and total flux ratios  from the models are listed in Table \ref{pos}, along with the results obtained from the image centroids where separating the two components is possible. 
Errors of the positions are estimated from the $\chi^2$ surfaces of each parameter where $\Delta\chi^2=1$, or  from the scatter in fits within each night, whichever is bigger. %Other model parameters are not reported in this work since we mostly focus on the positions of the center of light of each component. 

The models confirm that the smaller and more circular component, i.e., the donor, has higher surface brightness and total flux than the more elongated disk around the gainer. The ellipse size of the donor from the models, when averaged over all the epochs, is $0.62\pm0.16$ mas along the major axis, and $0.52\pm0.14$ mas along the minor axis, which confirmed the images that the donor is elongated, but slightly larger than that from the theoretical models \citep{Harmanec2002}. The averaged size of the disk surrounding the gainer is $1.04\pm0.11$ mas along the major axis, consistent with the size of the images of the first epoch as well as theoretical models \citep{Harmanec2002} and the H$\alpha$ disk of \citet{Schmitt2008}.   The minor axis of the disk  is $0.63\pm0.07$ mas, larger than that expected in  theoretical models \citep{Bisikalo2000, Linnell2000}, implying this extended structure is perhaps from the electron scattering and/or free-free emission from the halo above the poles of the gainer \citep{Jameson1976, Zeilik1982}.  

We also compared the flux ratios from our models with those obtained from $H$ band light curves. The light curves from literatures\footnote{\label{foot0}The light curve of \citet{Zeilik1982} at the $H$ band gives a $-0.61$ mag difference between the total magnitude of the system and the magnitude 
   at primary eclipse. Interpolating the $J$ and $K$ band light curves of \citet{Jameson1976} we can obtain an $H$ band magnitude difference of $-0.74$. Taking the   average of the two we get a difference of 0.675 mag, corresponding to a $\frac{f_1+f_2}{f_p}$ value of 1.86.}
 give a value of $\frac{f_1+f2}{f_p} = 1.86$, where $f_1, f_2$ and $f_p$ are the fluxes of the donor, the disk of the gainer and the flux at the primary eclipse. 
Because the donor is not completely eclipsed by the disk at primary eclipse \citep{Linnell2000},  $f_p = f_{2} + a f_{1}$, where $a$ is the fraction of the donor flux that goes through. Therefore  we can infer that $\frac{f_1}{f_2} > 0.86$, consistent with our flux ratios derived from 
   the models at phase 0.210 and 0.828, i.e., 1.16 and 1.32. In addition, taking the average of the two values, $\frac{f_1}{f_2}=1.24$, we get $a=0.165$,  suggesting that $16.5\%$ of flux from the donor goes through at the primary eclipse. This also implies that the donor contributes $\sim17\%$ of total flux at the primary eclipses, consistent with the 20$\%$ value of \citet{Linnell2000}. 

The goodness of fit of the models (included in each panel of Figure~\ref{images}) 
are in general similar to that of the images. Nevertheless, discrepancies exist between the models and the images. For instance, the components' separations  from the models are slightly smaller than the images. The gainer disk appears bigger in the models than in the images. 
 %In addition,  the flux ratios of the donor and the disk from the models at maximum elongation (2007Jul03, Jul04 and Jul12, see Table \ref{pos}) are very different from the other phases when the disk is partially blocked by the donor. 
 These properties demonstrate the complexity of $\beta$ Lyr and suggest that the disk may be asymmetric.
% due to radiation from the donor and the shock where the mass stream collides with the disk. 
%More particularly, the model disk in the first epoch (phase=0.035) is un-physically thin so that most of the background donor is not blocked. This may be because the disk is actually semi-transparent at the edges. So even though the donor is almost totally blocked by the disk, we can still see some light of the donor though the disk. 
They also highlight the deficiencies in our models and underscore the need for  a self-consistent model that treats the radiative transfer properly and accounts for all epochs simultaneously.

%===============================

%===================================================
\section{The orbit of \betlyr}
\label{orbit}
\begin{figure}[thb]
\begin{center}
{
\includegraphics[angle=90,width=3.3in]{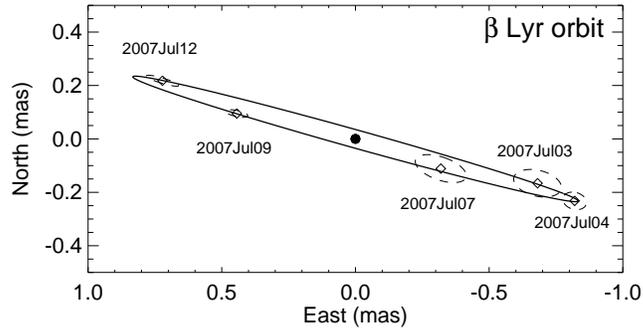}}
\hphantom{.....}
\caption{ % \scriptsize
The best fit relative orbit of \betlyr ({\it solid line}).  
The donor is indicated as a filled dot in the center. Positions of each epoch are shown by the open dots, surrounded by their error ellipses in dashed lines. The upper part of the ellipse is located towards the observer.
\label{orb}}
\end{center}
\end{figure}

Despite the deficiencies in our simplified models, the positions of each component's center of light are well-constrained, especially near the maximum elongation of the orbit.  Thus, the above results, along with the elements obtained from RV and light curve studies allow us to calculate the astrometric  orbit of $\beta$ Lyr for the first time. We adopt $P = 12.^{d}9414$ and T$_{mini} = JD2454283.0430$ (on 2007Jul01) from the recent ephemeris\footnote{\label{foot1}T$_{mini} $ (phase 0) is the epoch of primary minimum light.} \citep{Ak2007}, and $e=0$ \citep{Harmanec2002}. The best-fit orbit using the model positions is shown in Figure \ref{orb}. The resultant inclination, position angle of the ascending node ($\Omega$) and semi-major axis are listed in Table \ref{orbpar}. Other orbital solutions using positions from the images are also listed in the table.  Errors of  orbital elements are estimated using Monte-Carlo simulations. The 3 sets of inclination and $\Omega$ in Table \ref{orbpar} are consistent with each other and suggest a  retrograde orbit (i.e., position angle
decreasing with time). Our estimate on $\Omega$ is roughly consistent with the 248\fdg8 value in \citet{Schmitt2008}, and is almost perpendicular to the orientation of the jet (163\fdg5) implied by \citet{Hoffman1998}.

%The angular size of the orbit predicted by the images, however,  is larger than the size from the models, as expected from their different component separations. 
We can also estimate  the distance of $\beta$ Lyr using orbital parallax (see Table \ref{orbpar}) by combining its angular semi-major axis with the linear $a \sin i$ value, 57.87$\pm$0.62\rsun 
\footnote{\label{foot2}We obtain the semimajor axis by combining the
semiamplitude of the gainer $K_1$
($41.4 \pm 1.3$ km~s$^{-1}$, Harmanec \& Scholz 1993;
 $42.1 \pm 1.3$ km~s$^{-1}$ [error assumed], Bisikalo et al. 2000;
 $35.4 \pm 2.7$ km~s$^{-1}$ [from their Fig.~5], Ak et al. 2007;
 yielding a weighted average of $41.1 \pm 2.7$ km~s$^{-1}$)
with that for the donor $K_2$
($185.27 \pm 0.20$ km~s$^{-1}$, Ak et al. 2007). We derive $q=M_2/M_1=0.222\pm0.013, a\sin i=57.87\pm0.62$\rsun, $M_1\sin^3i=12.73\pm0.27$ \msun, and $M_2\sin^3i=2.82\pm0.18$ \msun}.  
The distance from our models,  314$\pm$17 pc, is larger than that from the images, 278$\pm$24 pc and 274$\pm$34 pc, but they are all consistent within errors with the $Hipparcos$ distance, 296$\pm$16 pc \citep{van-Leeuwen2007}. Finally, using the newly estimated $M \sin^3i$ for both components (see footnote \ref{foot2}) together with the inclination from the models,  we get mass of the gainer = 12.76$\pm$0.27\msun~and mass of the donor= 2.83$\pm$0.18\msun.

%======================================================

%========================================================

\section{Future work}
\label{summary}
%Our high resolution observations allowed us to obtain the first resolved images of $\beta$ Lyr. The images from the two methods consistently show the partially resolved mass donor and the thick disk surrounding the mass gainer at different epochs.  Our simple models interpret different epochs of observation well, providing orbital positions of the system and allowing us to improve the estimates of inclination and other orbital elements, which also allowed us to  estimate the distance of $\beta$ Lyr and the precise masses of the donor and the gainer.

We have only presented simple two-component models in this work since we mostly focus on the orbital positions of $\beta$ Lyr.   We have already discussed problematic discrepancies between the models and the images and also some internal inconsistencies between the model epochs.
The systematic difference in component separations between the images compared to the model fits
poses the most severe problem,  limiting the accuracy of our distance estimates to  $\sim15\%$.
To address these issues and better understand other physical properties of $\beta$ Lyrae, a more physical, self-consistent model is required that treats the radiative transfer and the sizes of the two components properly,  accounts for all epochs simultaneously, and incorporates the multi-wavelength information from eclipsing light curves.  
% is essential, and will be presented in a future work.

\acknowledgments
We thank Michael Ireland for the MACIM package used in this work. We also thank the referee's valuable suggestions and comments. The CHARA Array is funded by the National Science Foundation
through NSF grants AST-0307562 and AST-0606958 and by the Georgia State University.
We thank the support for this work by the Michelson Graduate Fellowship (MZ), the NSF Grants AST-0606861 (DG), NSF-AST 0352723, NSF-AST 0707927, NASA NNG 04GI33G (JDM), and EU grant MOIF-CT-2004-002990 (NT). EP was formally supported by the Michelson Postdoctoral Fellowship and is currently supported by  a Scottish
Universities Physics Association (SUPA) advanced fellowship.

\bibliographystyle{apj}  
\bibliography{betlyr}  

%\clearpage

\begin{deluxetable}{llccl}
\tabletypesize{\scriptsize}
%\rotate
\tablecaption{Observation logs for \betlyr}
\tablewidth{0pt}
\tablehead{
 \colhead{ Date (UT)} &\colhead{Mean MJD} & \colhead{Telescopes} & \colhead{N$_{blk}$} & \colhead{Calibrators}}
\startdata
 2006Oct16 & 54024.17  & W1-W2-S2-E2 &  1  & 29 Peg, $\upsilon$ And \\
2007Jul03   & 54284.25 & W1-W2-S1-E1 &   3 & $\gamma$ Lyr, $\upsilon$ Peg, $\upsilon$ And \\
 2007Jul04 & 54285.26   & W1-W2-S1-E1 &  3  & $\gamma$ Lyr, $\upsilon$ Peg, $\upsilon$ And \\
2007Jul07   &  54288.22 & W1-W2-S1-E1 &  3 & $\gamma$ Lyr, $\upsilon$ Peg, $\sigma$ Cyg\\
2007Jul09   & 54290.25 & W1-W2-S1-E1 &   3 & $\gamma$ Lyr, $\upsilon$ Peg\\
2007Jul12   & 54293.26 & W1-W2-S1-E1 &   3  & $\gamma$ Lyr, $\sigma$ Cyg
\enddata
\tablenotetext{a}{N$_{blk}$: number of data blocks}
\tablenotetext{b}{
Calibrator Diameters (milli-arcsec):  29 Peg = 1.017 $\pm$ 0.027, 
$\upsilon$ And = 1.098 $\pm$ 0.007 , $\sigma$ Cyg = 0.542 $\pm$ 0.021 (A. Merand 2008, private communication);  
$\gamma$ Lyr = 0.74 $\pm$0.10 \citep{Leggett1986}; $\upsilon$ Peg = 1.01 $\pm$ 0.04 \citep{Blackwell1994} }
%% Text for table notes should follow after the \enddata but before
%% the \end{deluxetable}. Make sure there is at least one \tablenotemark
%% in the table for each \tablenotetext.

\label{obslog}
\end{deluxetable}
%=====================================

\begin{deluxetable}{lllllllll}
\tabletypesize{\scriptsize}
%\rotate
\tablecaption{Orbital positions of  \betlyr}
\tablewidth{0pt}
\tablehead{
&  &\multicolumn{2}{c}{MACIM} & \multicolumn{2}{c}{BSMEM} & \multicolumn{3}{c}{Model} \\
   & & \colhead{Sep. } & \colhead{P.A.} & \colhead{Sep.} & \colhead{P.A.} &  \colhead{Sep.} & \colhead{P.A.} &  \colhead{Flux ratio}  \\
\multicolumn{1}{c}{Date}   & \colhead{Phase} & \colhead{(mas)} & \colhead{(deg)} & \colhead{(mas)} & \colhead{(deg)} &  \colhead{(mas)} & \colhead{(deg)}&  \colhead{(donor/gainer)} } 
 
\startdata
%UT 2006Oct16   & W1-W2-S2-E2  & 29 Peg, $\upsilon$ And \\
%2006Oct16    &0.035	&  --        & --            & --        &   --        &  0.156 $^{+ 0.200}_{-0.105}$ & 241.9 $\pm$ 24.5   & 0.62 $\pm$ 0.49 \\ %\tablenotemark{a}\\
2007Jul03    &0.132	&  0.811 &  255.4  & 0.853 & 253.7  &  0.701 $\pm$ 0.091  & 256.3 $\pm$ 4.0   & 1.01 $\pm$ 0.11 \\ 
2007Jul04    &0.210 & 0.891 &  253.3  & 0.886 & 254.4   &  0.852 $\pm$ 0.045 & 254.2 $\pm$ 2.1   & 1.16 $^{+ 0.20}_{-0.15}$  \\ 
2007Jul07    &0.438	& --        &   --         & --         & --          &   0.338 $\pm$ 0.105  & 250.8 $\pm$ 7.3   &  3.51 $\pm$ 1.27  \\ 
2007Jul09    &0.595	&  --       &   --         & 0.675 & 73.9  &   0.454 $\pm$ 0.042    & ~77.9 $\pm$ 1.4   & 2.43 $\pm$ 0.28  \\ 
2007Jul12    &0.828	& 0.842 &  72.3  & 0.783 & 69.6  &   0.754 $\pm$ 0.063      & ~73.2 $\pm$ 0.8   &  1.32 $^{+ 0.67}_{-0.27}$\\
\enddata
\tablenotetext{a}{Some positions are omitted for images whose centroids cannot be separated. }
\label{pos}
\end{deluxetable}

\clearpage

\begin{deluxetable}{lccc}
\tabletypesize{\scriptsize}
%\rotate
\tablecaption{Parameters of  \betlyr}
\tablewidth{0pt}
\tablehead{
\colhead{} & \colhead{MACIM}& \colhead{BSMEM}  & \colhead{Model} }
\startdata
%UT 2006Oct16   & W1-W2-S2-E2  & 29 Peg, $\upsilon$ And \\
Inclination (deg)  & 92.10 $\pm$ 1.24          &  91.96 $\pm$ 1.65   & 92.25   $\pm$ 0.82\\
$\Omega$ (deg)  & 253.22 $\pm$ 1.97        & 251.87 $\pm$ 1.83 & 254.39 $\pm$ 0.83 \\ 
semi-major axis (mas) & 0.976 $\pm$ 0.083 & 0.993 $\pm$ 0.122 & 0.865   $\pm$ 0.048 \\ 

\hline
Distance (pc)   & 276 $\pm$ 23   & 271 $\pm$ 33 & 312 $\pm$ 17
\enddata
\label{orbpar}
\end{deluxetable}

\clearpage

\end{document}